\documentstyle[epsfig,12pt]{article}
\title{{\bf Once more on extra quark-lepton generations and precision
measurements} \\ (dedicated to L.B. Okun's 80th birthday)}
\author{V.A. Novikov\thanks{e-mail: novikov@itep.ru} \\
ITEP, Moscow, Russia \\ A.N. Rozanov\thanks{e-mail: rozanov@cppm.in2p3.fr} \\
CPPM, Marseille, France; ITEP, Moscow, Russia \\
M.I. Vysotsky\thanks{e-mail: vysotsky@itep.ru}
\\ITEP, Moscow, Russia }
\date{}
\begin{document}
\maketitle

\begin{abstract}
Precision measurements of $Z$-boson parameters and $W$-boson and
$t$-quark masses put strong constraints on non $SU(2)\times U(1)$
singlet New Physics. We demonstrate that one extra generation
passes electroweak constraints even when all new particle masses
are well above their direct mass bounds.
\end{abstract}

\section{Introduction}

Nine years ago in paper \cite{1} it was noted that contrary to the
common belief expressed in review paper \cite{2} the precision
electroweak data do not exclude the existence of extra
quark-lepton generations. A year after in important paper \cite{3}
it was found that heavy Higgs boson does not contradict
electroweak data as soon as extra generation exists.

The purpose of the present paper is twofold. Firstly, to update old
results taking into account the present values of electroweak
observables. Secondly, the existence of
the fourth generation does not contradict precision data even if
the mass of new neutral lepton is much larger than $m_Z/2$
contrary to the statement
of the update of \cite{2} published in \cite{4}.

\section{Standard Model fit}

In Table 1 we present the results of the data fit performed by LEPTOP
code \cite{5}.

\begin{center}

{\bf Table 1}

\bigskip

\begin{tabular}{|l|l|l|r|}
\hline Observable & Exper. data  & LEPTOP fit & Pull \\ \hline
$\Gamma_Z$, GeV &    2.4952(23) &  2.4963(15)  & -0.5
\\ $\sigma_h$, nb &   41.540(37)& 41.476(14)  & 1.8    \\ $R_l$ &
20.771(25)& 20.743(18)  & 1.1   \\ $A_{\rm FB}^l$ & 0.0171(10)  &
0.0164(2)  & 0.8   \\ $A_{\tau}$ & 0.1439(43) &  0.1480(11)  &
-0.9   \\ $R_b$ &    0.2163(7) &   0.2158(1)  & 0.7   \\ $R_c$ &
0.172(3)&   0.1722(1)  & -0.0   \\ $A_{\rm FB}^b$  & 0.0992(16)&
0.1037(7)  & -2.8   \\ $A_{\rm FB}^c$  &    0.0707(35)& 0.0741(6)
& -1.0   \\ $s_l^2$ ($Q_{\rm FB}$)   &    0.2324(12)  & 0.2314(1)
& 0.8   \\ \hline
$A_{\rm LR}$ &  0.1513(21)& {0.1479(11)}  & 1.6
\\ $A_b$ &  0.923(20) & 0.9349(1)  & -0.6
\\ $A_c$ &  0.670(27)&   0.6682(5)  & 0.1   \\ \hline $m_W$, GeV &
80.398(25) & 80.377(17)  & 0.9   \\ \hline $m_t$, GeV    &
172.6(1.4) &   { 172.7(1.4)} &-0.1\\ $m_H$, GeV  &   &  {
$84^{+32}_{-24} $} &  \\ $\hat{\alpha}_s$ & &  0.1184(27) &  \\
$1/\bar{\alpha}$ & 128.954(48) & 128.940(46) & 0.3  \\ {\small
$\chi^2/n_{\rm d.o.f.}$} & & 18.1/12 & \\ \hline
\end{tabular}

\bigskip
Standard Model fit by LEPTOP, summer 2008.

\end{center}

The quality of the fit is characterized by $\chi^2/{\rm n.d.f.} =
18/12$ and is reasonably good. The central value of the Higgs
boson mass is well below LEP II direct search bound: $m_H \geq
114$ GeV at 95\% C.L.

\section{Fits with the fourth generation}

Introducing the fourth generation we get many new parameters:
quark and lepton masses and mixing with three existing
generations. To simplify the analysis let us suppose that mixing is
small.

In order to investigate 5-dimensional parameter space ($m_H, m_U,
m_D, m_E, m_N$) we use the results of \cite{6} where steep and
flat directions in the dependence of $\chi^2$ on new particle
masses were found. We fix the values of the sum of new quark
masses $m_U + m_D = 600$ GeV to avoid Tevatron direct search
bounds and the value of the charged lepton mass $m_E = 200$ GeV and
look for a minimum of $\chi^2$ for the fixed values of $m_H$,
varying neutral lepton mass and the difference of Up- and
Down-quark masses. The results of the fit are presented in Fig. 1
for $m_H = 120$ GeV, in Fig. 2 for $m_H = 600$ GeV
and in Fig. 3 for $m_H = 1000$ GeV. We see that
the quality of the fits is close
to that for the Standard Model without additional generation.


\begin{figure}[!htb]
\centering \epsfig{file=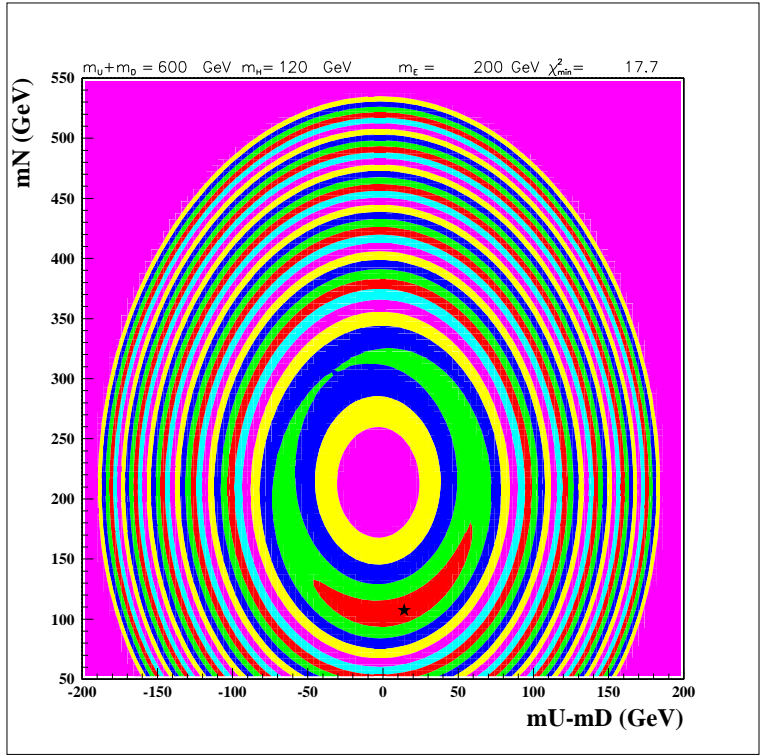,width=10cm} \caption{\em
Exclusion plot on the plane $m_N$, $m_U - m_D$ for fixed values
$m_H = 120$ GeV, $m_E = 200$ GeV, $m_U + m_D = 600$ GeV. $\chi^2$
minimum shown by the star corresponds to $\chi^2/d.o.f. =
17.7/11$.The borders of the regions show the domains allowed at the
level $\Delta\chi^2 = 1,4,9,16$ etc. } \label{WW2ermi}
\end{figure}


\begin{figure}[!htb]
\centering \epsfig{file=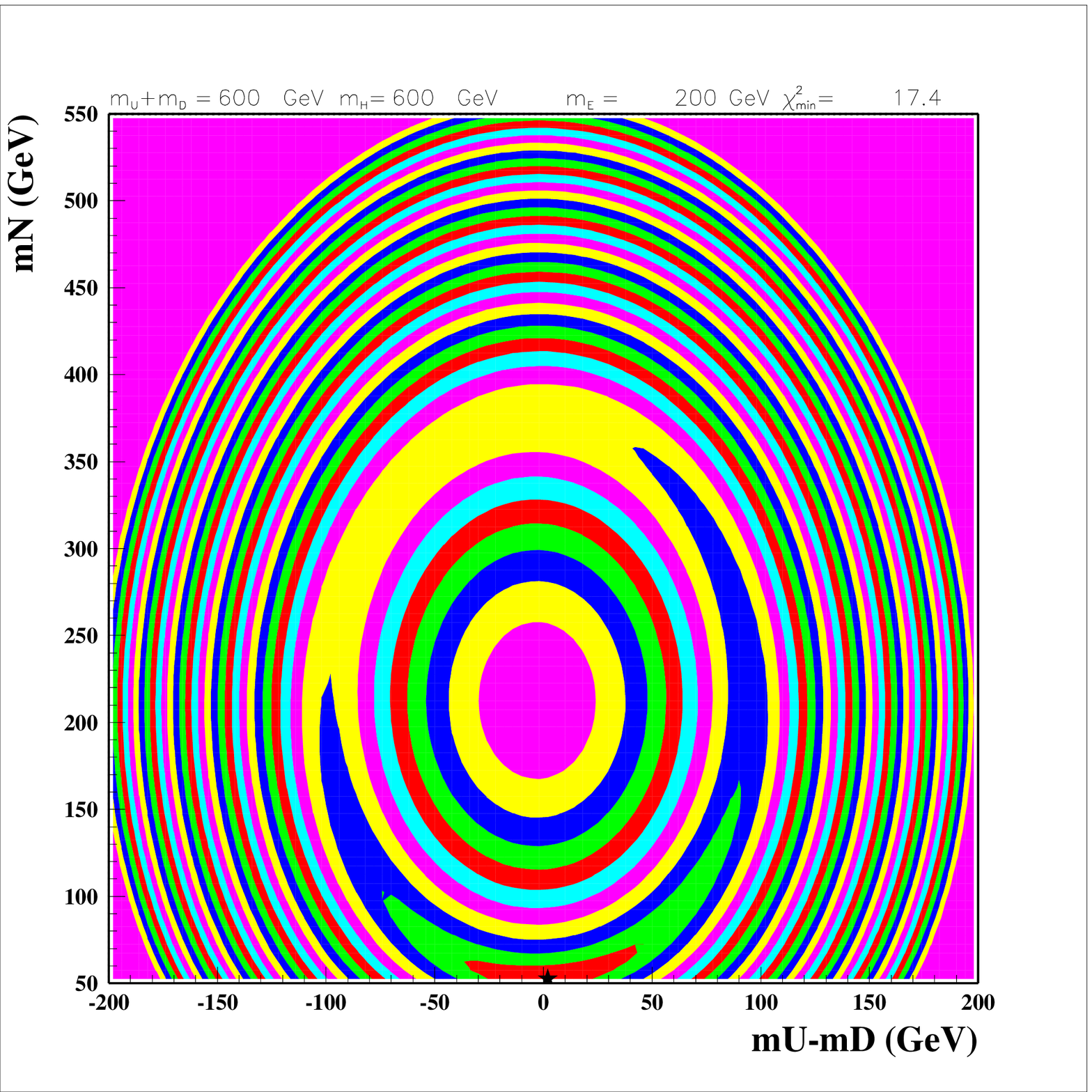,width=10cm} \caption{\em The same
as Fig. 1 for $m_H = 600$ GeV. $\chi^2$ minimum shown by the star
corresponds to $\chi^2/d.o.f. = 17.4/11$. } \label{W2Fermi}
\end{figure}

\begin{figure}[!htb]
\centering \epsfig{file=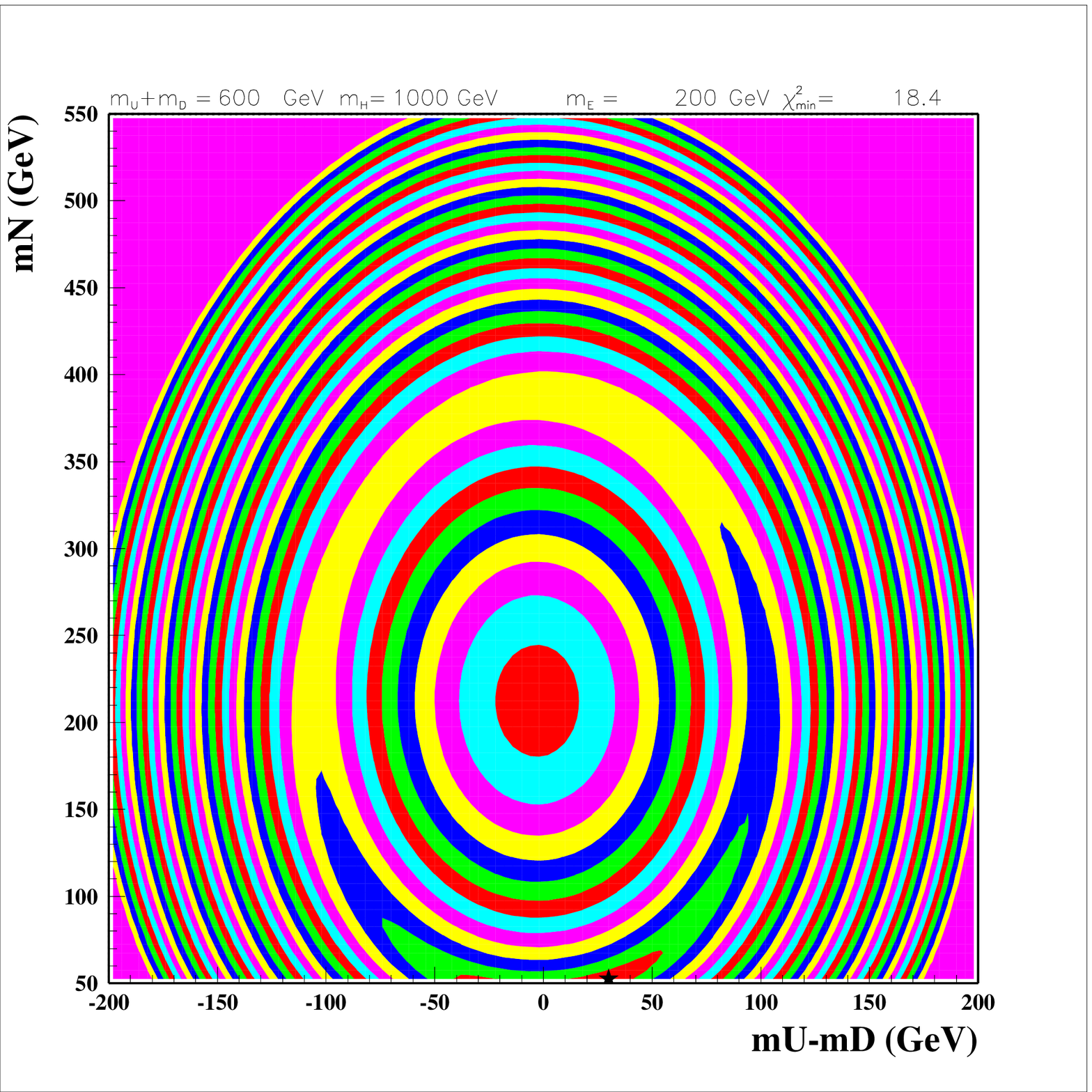,width=10cm} \caption{\em The same
as Fig. 1 for $m_H = 1000$ GeV. $\chi^2$ minimum shown by the star
corresponds to $\chi^2/d.o.f. = 18.4/11$. } \label{WW2Feri}
\end{figure}


\section{How many new generations?}

The next question which we address is the number of additional
generations allowed by precision data\footnote{To simplify the
analysis we assume the degeneracy of new particles with the identical
quantum numbers: $m_{E_1} = m_{E_2} = ...$, $m_{N_1} = m_{N_2} =
...$, $m_{U_1} = m_{U_2} = ...$, $m_{D_1} = m_{D_2} = ...$.}. To
study this problem we fix $m_E = 200$ GeV, $m_U = m_D = 300$ GeV,
since in  Figures 1, 2 and 3 minimum of $\chi^2$ corresponds to
almost degenerate quarks, and allow the number of extra generation
$N_g$ and mass of neutral lepton $m_N$ to vary, bounding Higgs
boson mass to be above direct search bound, $m_H > 114$ GeV. The
levels of $\chi^2$ are shown in Fig. 4.

\begin{figure}[!htb]
\centering \epsfig{file=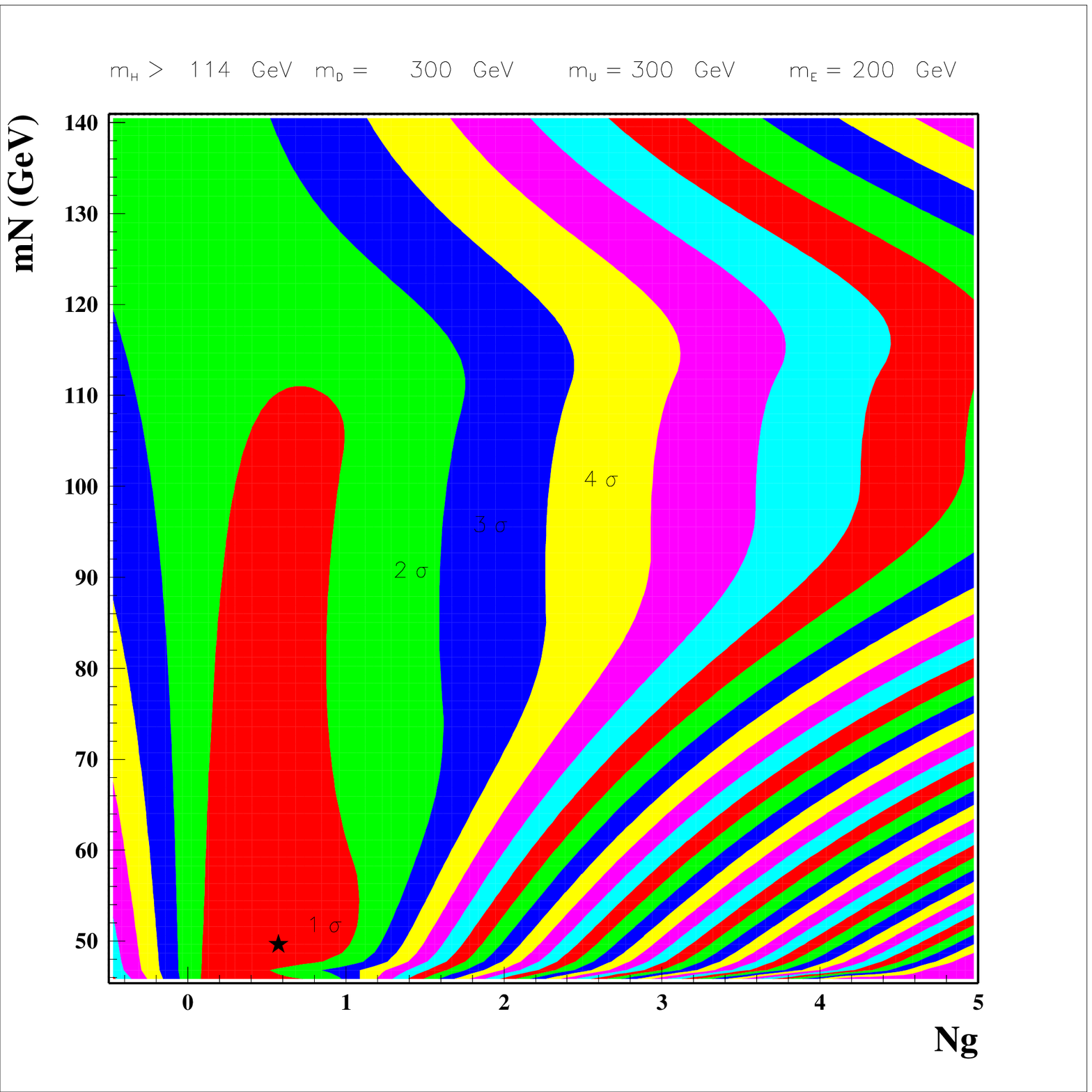,width=10cm} \caption{\em Exclusion
plot on the plane $N_g$, $m_N$ for fixed values $m_U = m_D = 300$
GeV, $m_E = 200$ GeV. $\chi^2$ minimum is shown by the star. The
condition $m_H
> 114$ GeV is imposed. } \label{WW3Fermi}
\end{figure}

As we already stressed the value of $\chi^2$ for Standard Model
and for $N_g = 1$ are almost the same, while three and more
additional generations are excluded.

In order to understand which masses of Higgs bosons correspond
to different
regions in Fig. 4 we draw Fig. 5. We see that for the
case $N_g = 1$ higgs mass rapidly grows when the neutral lepton mass
diminishes below 100 GeV.

\begin{figure}[!htb]
\centering \epsfig{file=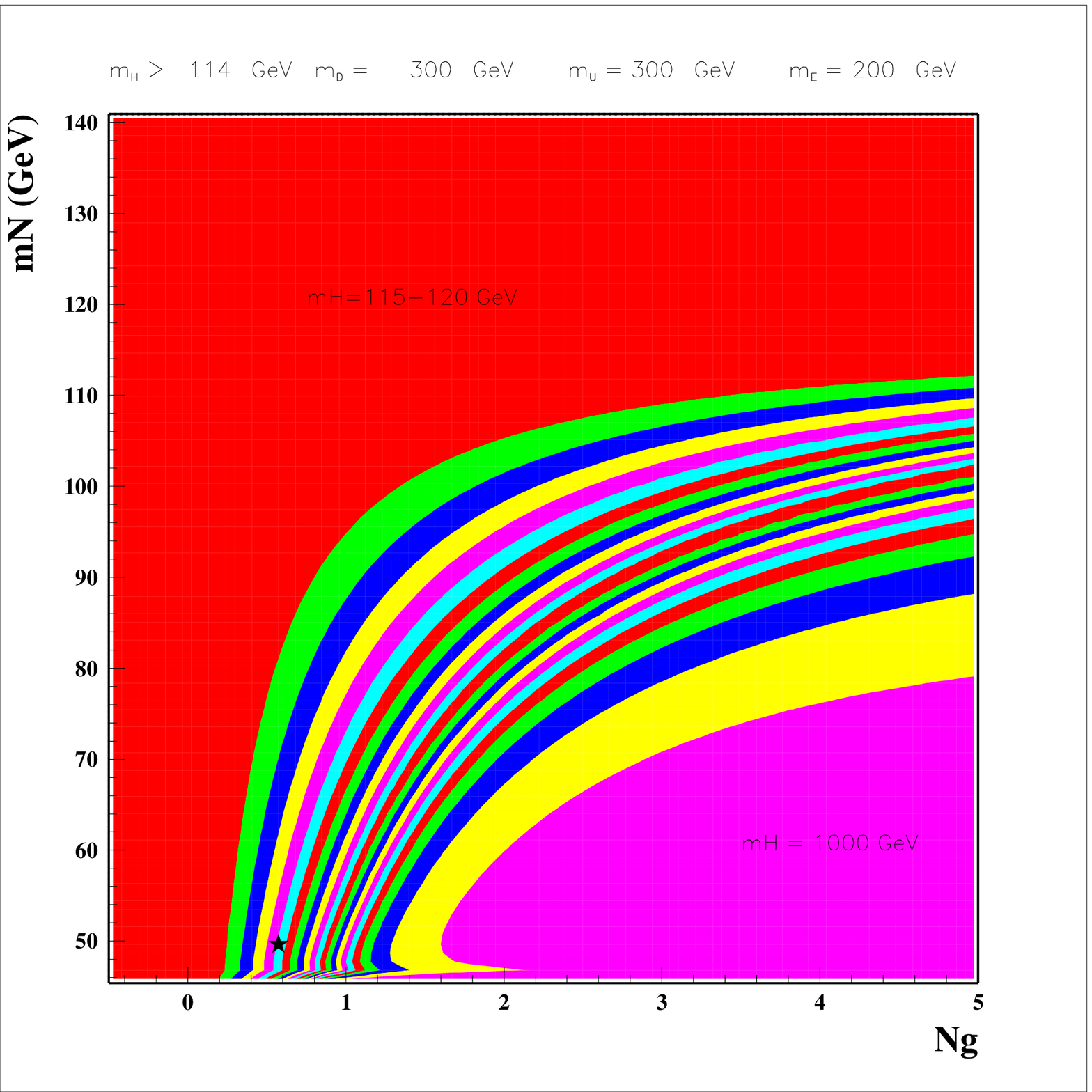,width=10cm} \caption{\em Values
of Higgs boson masses on the plot of Fig. 4 are shown. Each strip
corresponds to the definite value of $m_H$.} \label{WW4Fermi}
\end{figure}

\section{${\bf S, T, U}$ versus ${\bf V_m, V_A, V_R}$ -- formulas}

$S$, $T$ and $U$ variables were suggested in \cite{7} for the analysis
of New Physics contribution to electroweak observables. In paper
\cite{7} it was supposed that New Physics contributes only to the
vector bosons polarization operators and masses of new particles
are much heavier than $m_Z$, $m_W$. Extra generations not mixed
with three ``light'' ones satisfy the first requirement; concerning the
second one it depends on the masses of new particles. The expressions
for $S$, $T$ and $U$ introduced in \cite{7} take into account
the values of the polarization operators and their first derivatives at
$q^2 =0$. In this way the beautiful formulas were obtained with a clear
physical meaning: $S$ characterizes the violation of chiral symmetry,
while $T$ and $U$ characterize the isotopic symmetry violation.

Radiative corrections to electroweak observables were expressed in
LEPTOP through three functions $V_i$:
\begin{equation}
\frac{m_W}{m_Z} = c + \frac{3\bar\alpha c}{32\pi s^2(c^2 - s^2)}
V_m \;\; , \label{1}
\end{equation}
\begin{equation}
g_A = -\frac{1}{2} - \frac{3\bar\alpha}{64\pi c^2 s^2} V_A \;\; ,
\label{2}
\end{equation}
\begin{equation}
\frac{g_V}{g_A} = 1 - 4s^2 + \frac{3\bar\alpha}{4\pi(c^2 - s^2)}
V_R \;\; , \label{3}
\end{equation}
\begin{equation}
s^2c^2 \equiv \sin^2\theta_W \cos^2\theta_W =
\frac{\pi\bar\alpha}{\sqrt{2}G_\mu m_Z^2} \; , \;\; \bar\alpha
\equiv \alpha(m_Z) = (128.87)^{-1} \;\; , \label{4}
\end{equation}
where $g_A$ and $g_V$ are axial and vector couplings of $Z$-boson
with charged leptons. Functions $V_i$ depend both on Standard
Model and New Physics contributions. To make comparison with $S,
T, U$ approach let us separate these two contributions in $V_i$:
\begin{equation}
V_i \equiv V_i^{\rm SM} + \delta_{NP} V_i \;\; . \label{5}
\end{equation}

The definitions of $S$, $T$ and $U$ used in \cite{4} differ from the
original one used in papers \cite{7}: instead of the derivatives of
polarization operators at $q^2 = 0$ their values at $q^2 = m_W^2$
and $q^2 = m_Z^2$ were used. These new definitions coincide with
our functions $V_i$ with one exception: instead of derivative of
$Z$-boson polarization operator at $q^2 = m_Z^2$ which enters
function $V_A$, the difference $\Pi_Z(m_Z^2) - \Pi_Z(0)$ is used
in \cite{4}.

Comparing the definitions of $S$, $T$ and $U$ presented in Eqs.
(10.61a - 10.61c) from \cite{4} with definitions of our functions
$V_i$ presented in \cite{1} Eqs. (1) - (3) we get\footnote{Let us
note that our definition of polarization operators differs by sign
from that used in \cite{4}; also functions $V_R$ and $V_m$ contain
derivative of photon polarization operator at $q^2 = 0$, while $S$
and $U$ contain $\Pi_\gamma(m_Z^2)/m_Z^2$.}:
\begin{equation}
T = \frac{3}{16\pi s^2 c^2} \delta_{NP} V_A + \Delta \equiv
T^\prime + \Delta \;\; , \label{6}
\end{equation}
\begin{equation}
S = \frac{3}{4\pi}[\delta_{NP} V_A - \delta_{NP} V_R] + 4s^2 c^2
\Delta \equiv S^\prime + 4s^2 c^2 \Delta \;\; , \label{7}
\end{equation}
\begin{equation}
S+U = \frac{3}{4\pi(c^2 - s^2)} (\delta_{NP} V_m - \delta_{NP}
V_R) \equiv S^\prime + U^\prime \;\; , \label{8}
\end{equation}
\begin{equation}
\Delta \equiv \frac{1}{\bar\alpha}\left[\Pi_Z^\prime(m_Z^2) -
\frac{\Pi_Z(m_Z^2)}{m_Z^2} + \frac{\Pi_Z(0)}{m_Z^2}\right] \;\; ,
\label{9}
\end{equation}
where we introduce new set of functions $S^\prime$, $T^\prime$ and
$U^\prime$ which are directly related to $\delta_{NP}V_i$. If the
authors of \cite{4} use these primed functions in the expressions
for physical observables instead of $S$, $T$ and $U$, then their
formulas would be exact for arbitrary values of the new particles
masses. The accuracy of the formulas for physical observables through
$S$, $T$ and $U$ is good as long as new particles are considerably
heavier than $Z$-boson, since in this case $\Delta$ is suppressed
as $(m_Z/2m)^2$ (we substituted $2m$ instead of $m$ since cut of
$Z$-boson polarization operator starts at $q^2 = (2m)^2$). The
authors of \cite{4} recognize that for the light fourth generation
particles their approach based on $S$, $T$ and $U$ can be wrong.

Our next goal is to present explicit formulas for the fourth
generation contributions to $T$, $S$ and $\Delta$. Let us start
from the expression for $\delta_{NP}V_A$:
\begin{equation}
\delta_{NP}V_A = \frac{16\pi s^2
c^2}{3\bar\alpha}\left[\frac{\Pi_Z(m_Z^2)}{m_Z^2} -
\frac{\Pi_W(0)}{m_W^2} - \Pi_Z^\prime(m_Z^2)\right] \label{10}
\end{equation}
and compare it with that for $T$ from \cite{4}, Eqs. (10.61a,
10.63, 10.59):
\begin{eqnarray}
\bar\alpha T & = & \frac{\Pi_Z(0)}{m_Z^2} - \frac{\Pi_W(0)}{m_W^2}
= \frac{3\bar\alpha}{16\pi c^2 s^2} \left[u + d - \frac{2ud}{u-d}
\ln\left(\frac{u}{d}\right)\right] + \nonumber \\
& + & \frac{\bar\alpha}{16\pi c^2 s^2} \left[N + E -
\frac{2NE}{N-E} \ln \left(\frac{N}{E}\right)\right] \;\; ,
\label{11}
\end{eqnarray}
where $u \equiv m_U^2/m_Z^2$, $d \equiv m_D^2/m_Z^2$, $N \equiv
m_N^2/m_Z^2$ and $E\equiv m_E^2/m_Z^2$. With the help of
Eq.(\ref{6}) we obtain:
\begin{eqnarray}
\Delta & = & - \frac{3}{16\pi s^2 c^2} \delta_{NP} V_A +
\frac{1}{\bar\alpha} \left[\frac{\Pi_Z(0)}{m_Z^2} -
\frac{\Pi_W(0)}{m_W^2}\right] =
\nonumber \\
& = & \frac{3}{16\pi s^2
c^2}\left\{\frac{N_c^{q,l}}{3}\left[F^\prime(u) +
F^\prime(d)\right] + N_c^{q,l}\left(\frac{4}{9} s^2
+\frac{1}{9}\right)  \times \right. \nonumber \\
& \times & \left[2u F(u) - (1+ 2u) F^\prime(u) + 2d F(d) - (1+ 2d)
F^\prime(d)\right] + \nonumber \\
& + & \frac{16}{9} N_c^{q,l} s^4
\left[Q_u^2\left((1+2u)F^\prime(u) -2u F(u)\right) + \right. \\
& + & \left.  Q_d^2 ((1+2d) F^\prime(d) - 2d
F(d))\right] + \frac{4}{9} N_c^{q,l} s^2 Y^{q,l} \times
\nonumber \\
& \times & \left. \left[(1+2d) F^\prime(d) - 2d F(d) + 2u F(u) -
(1+2u) F^\prime(u)\right] \right\} \nonumber \;\; , \label{12}
\end{eqnarray}
where the contributions of quarks and leptons should be summed up,
$N_c^q = 3$, $N_c^l = 1$, $Y_q = Q_u + Q_d = 1/3$, $Y_l = Q_N +
Q_E = -1$,
\begin{equation}
F(u) = 2\left[1-\sqrt{4u-1} \arcsin \frac{1}{\sqrt{4u}}\right]
\;\; , \label{13}
\end{equation}
$$ F^\prime(u) \equiv -u\frac{d}{du} F(u) = \frac{1-2u F(u)}{4u-1}
\;\; , $$ see Eqs. (5), (9), (10) from \cite{1}.

From Eqs. (7) and (12) we get:
\begin{equation}
S = -\frac{3}{4\pi} \delta_{NP} V_R + \frac{4s^2 c^2}{\bar\alpha}
\left[\frac{\Pi_Z(0)}{m_Z^2} - \frac{\Pi_W(0)}{m_W^2}\right] \;\;
, \label{14}
\end{equation}
and with the help of Eq. (11) and Eq. (6) from \cite{1} we finally
get:
\begin{eqnarray}
S & = & \frac{3}{4\pi}\left\{\frac{2N_c^{q,l}}{3} [uF(u) + dF(d)]
- \frac{16}{9} N_c^{q,l} s^2 c^2 \left[Q_u^2\left((1+2u) F(u) -
\frac{1}{3}\right)\right. \right. + \nonumber \\
& + & \left. Q_d^2\left((1+2d) F(d) - \frac{1}{3}\right)\right] -
\frac{2N_c^{q,l} Y^{q,l}}{9} \left[(1+
2d) F(d) - \right. \nonumber \\
& - & \left.\left. (1+2u)F(u) +
\ln\left(\frac{u}{d}\right)\right]\right\} \;\; , \label{15}
\end{eqnarray}
where contributions of quarks and leptons should be summed up.

\section{${\bf S, T, U}$ versus ${\bf V_m, V_A, V_R}$ -- numbers}

In Section 3 we found three points of $\chi^2$ minimum in the fourth
generation parameter space, which corresponds to $m_H = 120$ GeV,
$m_H = 600$ GeV and $m_H = 1000$ GeV. The values of quark and lepton contributions
to $S$, $S^\prime$, $T$, $T^\prime$, $U$ and $U^\prime$ at these points are
presented in Table 2.

\newpage
\begin{center}

{\bf Table 2}

\bigskip

\begin{tabular}{|c|c|c|c|c|c|c|}
\hline & \multicolumn{2}{c|} {$m_H = 120 $} & \multicolumn{2}{c|}{$m_H = 600$} & \multicolumn{2}{c|}{$m_H = 1000$}\\
\hline & $m_U$ =  310 & $m_N = 120 $ & $m_U = 300$ & $m_N = 50 $& $m_u=315$&$m_N=53$\\
& $m_D = 290 $ & $m_E = 200$ & $m_D = 300 $ & $m_E = 200 $&$m_D=285$&$m_E=200$ \\
\hline $T^\prime$ & 0.02 & 0.11 & 0 & 0.24 &0.05& 0.27\\ \hline
$T$ & 0.02 & 0.11 & 0 & 0.37&0.05&0.36
\\ \hline $S^\prime$ &
0.15 & -0.01 & 0.16 & -0.23&0.15&-0.19 \\ \hline $S$ & 0.15 & -0.01 & 0.16 & -0.14 &0.15&-0.13\\
\hline $U^\prime$&0&0.02&0&0.20&0&0.16\\ \hline
$U$&0&0.01&0&0.11&0&0.10\\\hline
\end{tabular}

\bigskip

{\em Quark and lepton contributions to $S, T, U$ and 
$S^\prime, T^\prime, U^\prime $ at the points of $\chi^2$
minimum in Figures 1,2  and 3. All masses are in GeV.}

\end{center}

For heavy higgs  the mass of a neutral lepton is close
to $m_Z/2$, that is why the values of $S^\prime$, $T^\prime$ 
and $U^\prime$
considerably differ from $S$, $T$ and $U$. For light higgs ($m_H =
120$ GeV) the masses of all new fermions are far from $m_Z/2$, that is
why $S$, $T$ and $U$ almost coincide with $S^\prime$, $T^\prime$
and $U^\prime$.

In order to get whether these sets of the fourth generation particle
masses are allowed by precision data in the framework of $S$, $T$,
$U$ approach, we should look at Fig. 10.4 from \cite{4}. Standard
Model corresponds to $S = T = 0$, and this point is just at the
border of 90 \% C.L. allowed domain for $m_H = 117$ GeV.

Summing quarks and leptons contributions for the fourth generation
point with $m_H = 120$ GeV we get $S = 0.14$, $T = 0.13$. It is at
the border of the same domain, so its level of $\chi^2$ is the
same as in the case of SM -- the result coinciding with that of
$V_i$ analysis presented in Section 3.

Next we should consider the point with heavy higgs, $m_H = 1000$
GeV. At this point $S = 0.02$, $T = 0.41$, which is a bit outside the
allowed domain for 1000 GeV higgs from Fig. 10.4. However since at
this point $S^\prime = -0.04$, $T^\prime = 0.32$ we see that
primed quantities are again at the border of 90 \% C.L. allowed
domain.

\section{Conclusions}

The possibility to include an extra quark-lepton generation(s)
into Standard Model has been studied. We have found that
the electroweak data do
not contradict the existence of one extra family with specially
adjusted masses. Three examples corresponding to light and heavy
higgs bosons are presented. The properly made analysis
based on $S$, $T$, $U$ (for $m_H = 120$ GeV) and $S^\prime$,
$T^\prime$, $U^\prime$ (for $m_H = 1000$ GeV) confirms the results
of the analysis based on $V_i$. Let us note that in paper \cite{9}
a set of masses of the fourth generation particles was found which
pass electroweak tests and corresponds to $m_H = 115 - 300$ GeV.

This paper is an expanded version of the presentation at
the ``Beyond the 3 SM generation at the LHC era'' workshop at CERN, September 4-5 2008. We dedicate it to Lev Okun, who initiated our common
electroweak project many years ago.

This work was partly supported
by Rosatom; V.N. and M.V. were supported by grants RFBR 07-02-00021,
08-02-00494 and NSh-4568.2008.2; V.N. was supported also by grant
RFBR 07-02-00830.

\end{document}